\documentclass[twocolumn,prd,footinbib,aps,prl,floats,floatfix,amsmath,superscriptaddress,
amssymb,longbibliography,secnumarabic]{revtex4-1} %
\usepackage[final]{graphicx}
\usepackage{hyperref}
\usepackage{amsmath}
\usepackage{bbm}
\usepackage{amsfonts}
\usepackage{amssymb}
\usepackage{latexsym}
\usepackage{graphicx}
\graphicspath{{./figs/}}
\usepackage[english]{babel}
\usepackage{multirow}
\usepackage{float}
\usepackage{url}
\usepackage{slashed}
\usepackage{xcolor} 
\usepackage[utf8]{inputenc}
\usepackage{verbatim}
\usepackage{stackengine}
\usepackage{gensymb}
\usepackage[normalem]{ulem}

\def\sfrac#1#2{{\textstyle{#1\over #2}}}
\newcommand{\be}{\begin{equation}}
\newcommand{\ee}{\end{equation}}
\newcommand{\ba}{\begin{array}}
\newcommand{\ea}{\end{array}}
\newcommand{\bea}{\begin{eqnarray}}
\newcommand{\eea}{\end{eqnarray}}
\newcommand{\sss}{\scriptscriptstyle}

\newcommand{\nn}{\nonumber}

\newcommand{\nova}{NO$\nu$A\ }

\usepackage[utf8]{inputenc}

\begin{document}
\rightline{CERN-TH-2024-114}
\title{Dark photon distortions of NO$\nu$A and T2K neutrino oscillations}

\author{Gonzalo Alonso-{\'A}lvarez}
\email{gonzalo.alonso@utoronto.ca}
\thanks{ORCID: \href{https://orcid.org/0000-0002-5206-1177}{0000-0002-5206-1177}}
\affiliation{Department of Physics, University of Toronto, Toronto, ON M5S 1A7, Canada}
\author{James M.\ Cline}
\email{jcline@physics.mcgill.ca}
\thanks{ORCID: \href{https://orcid.org/0000-0001-7437-4193}{0000-0001-7437-4193}}
\affiliation{McGill University Department of Physics \& Trottier Space Institute, 3600 Rue University, Montr\'eal, QC, H3A 2T8, Canada}
\affiliation{CERN, Theoretical Physics Department, Geneva, Switzerland}
\author{Benoit Laurent}
\email{benoit.laurent@mail.mcgill.ca}
\thanks{ORCID: \href{https://orcid.org/0000-0002-1306-3620}{0000-0002-1306-3620}}
\affiliation{McGill University Department of Physics \& Trottier Space Institute, 3600 Rue University, Montr\'eal, QC, H3A 2T8, Canada}
\author{Ushak Rahaman}
\email{ushak.rahaman@cern.ch}
\affiliation{IFIC - Instituto de Física Corpuscular (CSIC - Universitat de València), c/Catedrático José Beltrán, 2, 46980 Paterna, Valencia, Spain}
\affiliation{Department of Physics, University of Toronto, Toronto, ON M5S 1A7, Canada}

\begin{abstract}
Dark photons coupling to $L_\mu-L_\tau$ lepton number difference are a highly studied light dark matter candidate, with potential to be discovered through their impact on terrestrial neutrino oscillation experiments.  We re-examine this in the light of claimed tensions between the 
NO$\nu$A and T2K long baseline experiments, also taking into account data from the MINOS experiment.  We obtain leading limits on the $L_\mu-L_\tau$ gauge coupling $g'$ versus dark photon mass $m_{A'}$, improving upon previous bounds by up to a factor of five. We find no statistically significant alleviation of the tension from inclusion of the new physics effect.  %We confirm that the combined data mildly prefer the inverted neutrino mass ordering.
\end{abstract}

\maketitle

\section{Introduction}
\label{sec:introduction}

Dark photons are a popular light dark matter candidate, which do not suffer from the naturalness problems
typical of light scalars, since the Stueckelberg mass is not radiatively corrected.  
If the dark photon is a gauge boson, it can couple to one of the anomaly-free currents present in the Standard Model (SM) without the need to add extra matter.
Here we focus on the lepton number family difference $L_\mu-L_\tau$, but similar phenomenology arises for $L_e-L_\mu$, $L_e-L_\tau$, and combinations of the three, though couplings to electrons are most strongly constrained by other processes.  
For the range of gauge boson masses of interest, very small values of the gauge coupling $g'$ can be probed through cosmology \cite{Huang:2017egl,Escudero:2019gzq,Dror:2020fbh,Hannestad:2004qu,Hannestad:2005ex,Escudero:2019gfk,Escudero:2020ped,Barenboim:2020vrr,Chen:2022idm}, neutron star binaries \cite{KumarPoddar:2019ceq, Dror:2019uea}, supernova SN1987A~\cite{Farzan:2002wx,Croon:2020lrf}, black hole superradiance~\cite{Baryakhtar:2017ngi,Cardoso:2017kgn,Cardoso:2018tly}, and even the LHC \cite{Ekhterachian:2021rkx}.

The effect of such dark photons on neutrino oscillations gives the strongest limits on the gauge coupling for $m_{A'}\lesssim 10^{-10}\,$eV \cite{Brdar:2017kbt,Brzeminski:2022rkf,Alonso-Alvarez:2021pgy,Alonso-Alvarez:2023tii,Lin:2023xyk}, assuming that the
polarization of $\vec A'$ is coherent over the neutrino baseline.
This is a reasonable assumption as long as the correlation length of the field, $(m_{A'}\sigma)^{-1}$, where $\sigma$ is the local dark matter velocity dispersion, is larger than the neutrino oscillation baseline. For the long-baseline neutrino oscillation experiments under study here, this imposes the requirement $m_{A'}\lesssim 10^{-10}$~eV. 
For larger masses, the polarization varies during the neutrino time of flight, the term linear in $A'$ averages out and the leading order term is $\mathcal{O}(g'A')^2$, which can only affect neutrino oscillations for gauge couplings already excluded by other constraints~\cite{Lin:2023xyk}.

The gauge interaction enters the effective neutrino Hamiltonian linearly,
\bea
    &H&=\frac{1}{2E}U\left[
\begin{array}{ccc}
m_{1}^{2} & 0 & 0\\
0 & m_{2}^{2} & 0\\
0 & 0 & m_{3}^{2}\\
\end{array}
\right]U^\dagger+\sqrt{2}G_FN_e \left[
\begin{array}{ccc}
1 & 0 & 0\\
0 & 0 & 0\\
0 & 0 & 0\\
\end{array}
\right]\nn\\
&+&\!\!g^\prime A^\prime\cos\!\phi\,\cos(m_{A^\prime}t+\alpha)\!\left[
\begin{array}{ccc}
q_e& 0 & 0\\
0 & q_\mu & 0\\
0 & 0 & q_\tau\\
\end{array}
\right]\! +\! {\cal O}(g'^2A'^2),\qquad
\label{Heq}
\eea
written in the $\nu$ flavor basis.
Here, $E$ is the neutrino energy, $U$ is the PMNS matrix, $N_e$ is the electron density in the Earth, $m_{A'}$ is the dark photon mass, $\alpha$ is the dark photon's phase when the neutrino is produced (which we assume to happen at $t=0$), and $\phi$ is the angle between $\vec A'$ and the neutrino momentum $\vec p$. The charge assignments are restricted to satisfy $q_e + q_\mu + q_\tau = 0$, as any diagonal contribution only introduces a global phase unobservable in oscillation experiments.

As mentioned above, we focus on the $L_\mu-L_\tau$ dark photon, and thus fix $q_e=0$, $q_\mu=1$, and $q_\tau=-1$.
Lepton number family differences involving electrons are much more strongly constrained by limits arising from tests of the equivalence principle~\cite{Schlamminger:2007ht}, fifth-force searches~\cite{Kapner:2006si,Salumbides:2013dua}, and matter effects on neutrino oscillations mediated by the dark photon~\cite{Wise:2018rnb,Coloma:2020gfv}.

An important feature of dark photon dark matter is the existence of a polarization.
Many dark photon dark matter cosmological production mechanisms~\cite{Nelson:2011sf,Arias:2012az,Graham:2015rva,Agrawal:2018vin,Co:2018lka,Bastero-Gil:2018uel,Alonso-Alvarez:2019ixv,Ema:2019yrd,Long:2019lwl,Ahmed:2020fhc,Kolb:2020fwh,Bastero-Gil:2021wsf} predict the existence of a primordial polarization; however, its evolution through nonlinear structure formation is an open question~\cite{Caputo:2021eaa,Gorghetto:2022sue,Amaral:2024tjg,Chen:2024vgh}.
In many models, the vectors are produced in arbitrary directions even if one of the polarizations is dominantly sourced~\cite{Graham:2015rva,Agrawal:2018vin,Co:2018lka,Bastero-Gil:2018uel,Ema:2019yrd,Long:2019lwl,Ahmed:2020fhc,Kolb:2020fwh,Bastero-Gil:2021wsf}.
If that is the case, it has been argued~\cite{Amaral:2024tjg} that virialization during gravitational clustering leads to an equipartition of polarizations.
In comparison, misalignment-like scenarios~\cite{Nelson:2011sf,Arias:2012az,Alonso-Alvarez:2019ixv} produce vectors oscillating in a single direction, which would be largely preserved through structure formation.

In what follows, we take the dark photon polarization to be fixed.
This is certainly the case in the single direction scenario, in which $A'$ can be taken to be a fixed vector in space throughout the whole experimental data taking period, with $\cos\phi$ only varying due to the Earth's motion.
In the equipartition scenario, the $A'$ field traces out an ellipse whose orientation is randomized after a coherence time $(m_{A'}\sigma)^{-1}$.
As long as the orientation persists during the neutrino time of flight, this still leads to a nonzero effect at the linear field level that we study here, with extra secular variation at the coherence time scale.
Since one motivation for this work is to investigate whether the tension between NO$\nu$A and T2K can be alleviated (as has been claimed in Ref.\ \cite{Lin:2023xyk}), we focus on the most favorable scenario of a fixed polarization, and leave the study of the secularly varying polarization for the future.

Our study is partially motivated by a mild tension between the T2K \cite{T2K:2023smv} and \nova \cite{NOvA:2021nfi,NOvA:2023iam} long baseline experiments, present already in their first data \cite{Nizam:2018got}. The tension grew stronger with accumulated data, reaching a complete mismatch between \nova and T2K allowed regions on $\sin^2 \theta_{23}-\delta_{\rm CP}$ plane, which attracted attention \cite{Kelly:2020fkv,Chatterjee:2020kkm,Cherchiglia:2023ojf, Rahaman:2021zzm, Rahaman:2022rfp}. Under the assumption of normal ordering of the neutrino masses, the preferred regions for the CP-violating Dirac phases are at odds between the two experiments, at the 2-$\sigma$ level.  The tension between \nova and T2K persists in the new result published by \nova collaboration after 10 years of data taking in Neutrino 2024 \cite{Wolcott:2024}.  Recently, Ref.\ \cite{Lin:2023xyk} investigated the effect of dark photons in this context, and found the tension to be alleviated by $\Delta\chi^2 = 4.8$, a 1.7$\sigma$ improvement.

\begin{figure*}[t]
\centerline{\includegraphics[width=0.8\columnwidth]{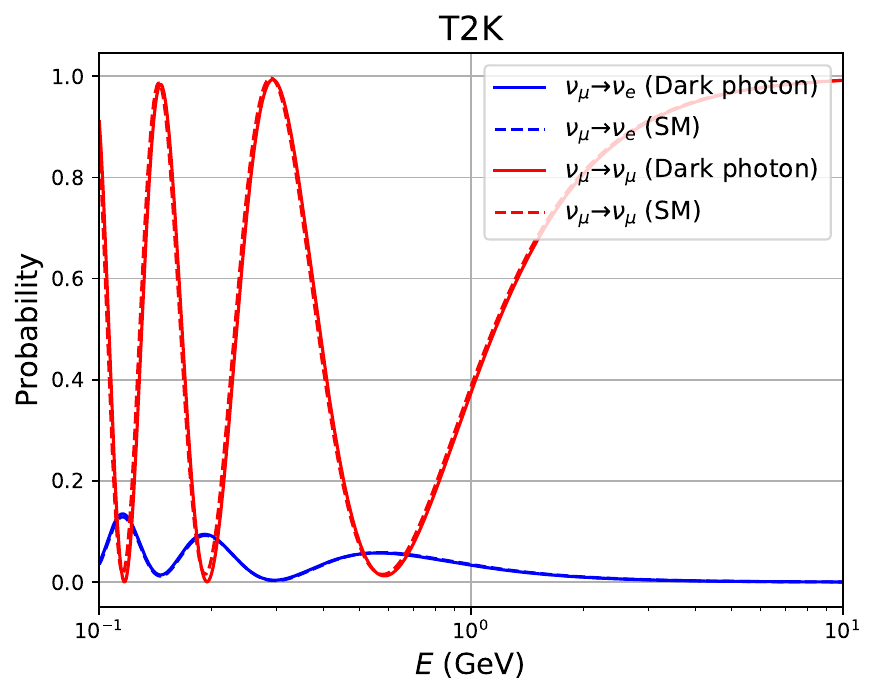}
\includegraphics[width=0.8\columnwidth]{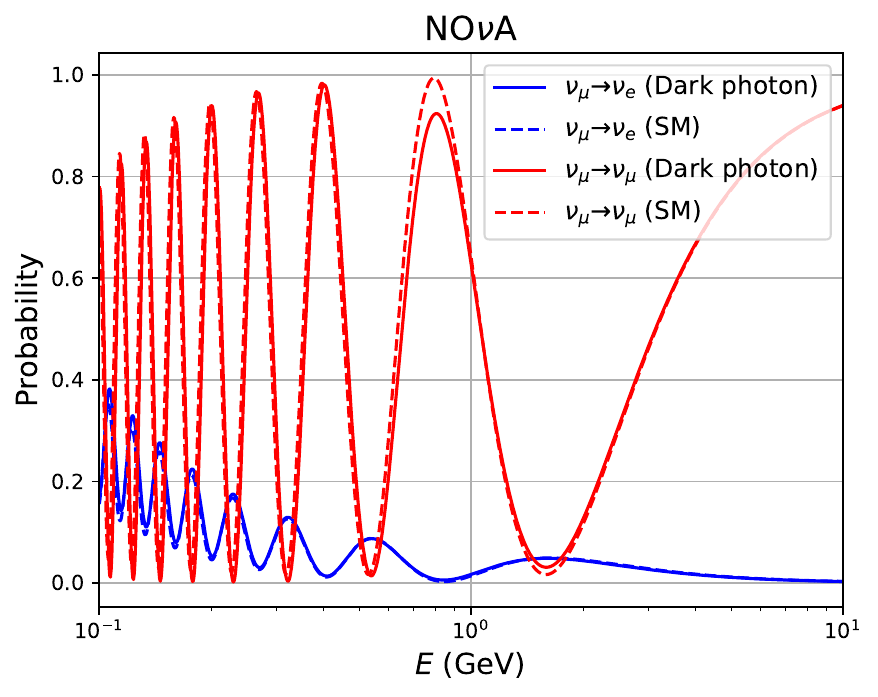}}
\centerline{\includegraphics[width=0.8\columnwidth]{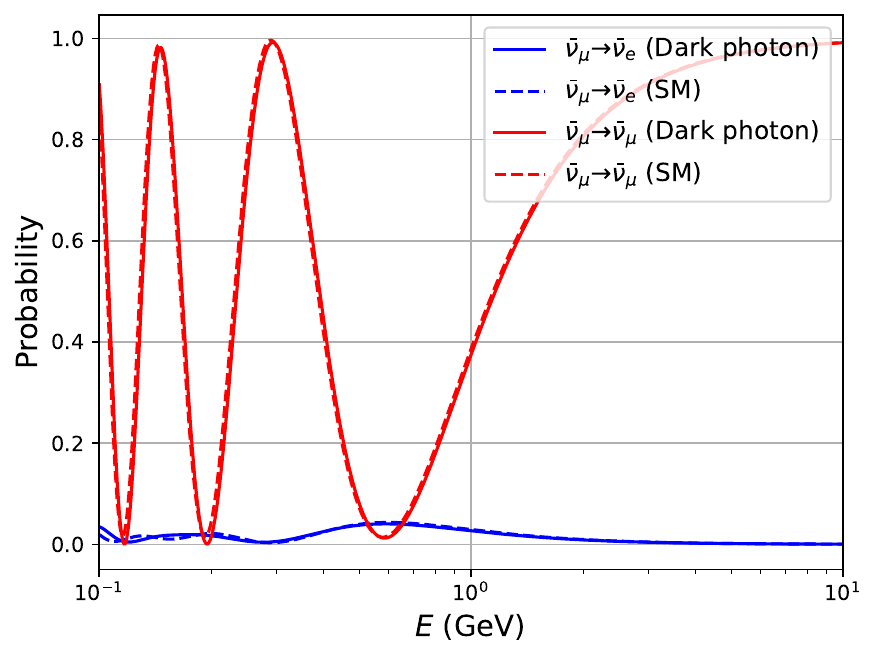}
\includegraphics[width=0.8\columnwidth]{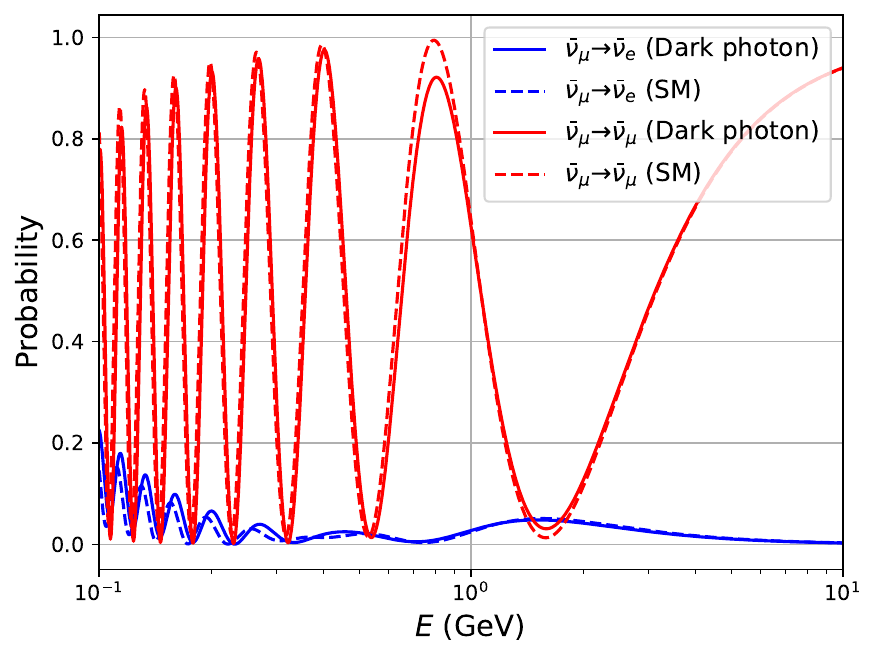}}
%\vspace{}
\caption{$\nu_\mu\to \nu_e$ (blue lines) and $\nu_\mu \to \nu_\mu$ (red lines) probabilities for T2K (NO$\nu$A) in the left (right) panels. The top (bottom) panels represent probabilities for neutrino (anti-neutrino) beam. The standard and non-standard parameter values have been fixed at the best-fit values of the combined analysis of NO$\nu$A, T2K, and MINOS.
}
\label{fig:prob}
\end{figure*}

\begin{figure*}[t]
\centerline{\includegraphics[width=2 \columnwidth]{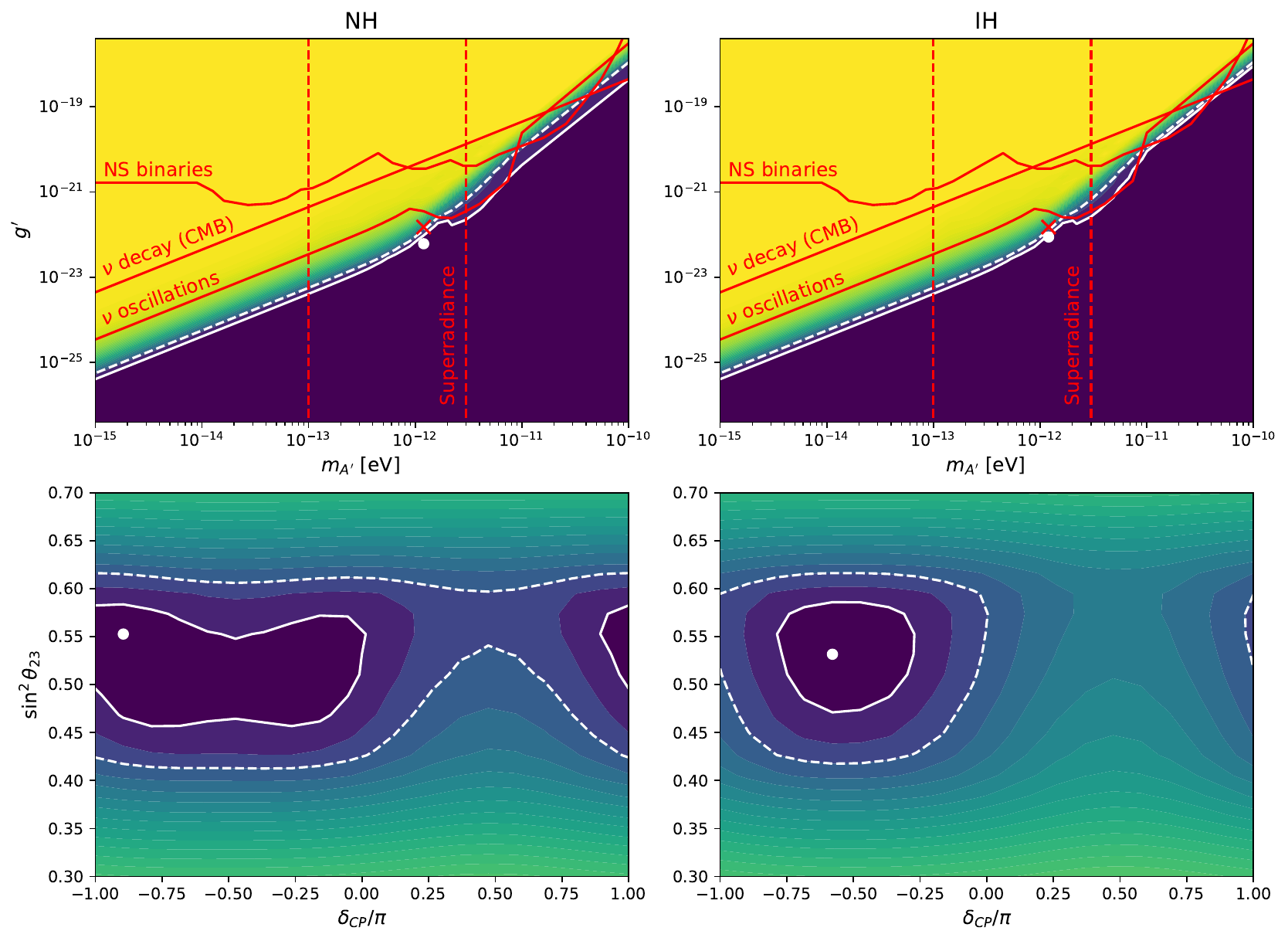}}
%\vspace{}
\caption{Top: upper bound on gauge coupling $g'$ versus dark photon mass $m_{A'}$ assuming 
the normal mass hierarchy (NH, left) and inverted hierarchy (IH, right), with best fit point shown in white. Each color shade corresponds to a difference of 1-$\sigma$, and the solid and dashed white lines highlight the 1-$\sigma$ and 3-$\sigma$ regions, respectively.
Previous limits from neutron star (NS) binaries~\cite{KumarPoddar:2019ceq, Dror:2019uea}, cosmology~\cite{Chen:2022idm}, neutrino oscillations~\cite{Alonso-Alvarez:2023tii}, and superradiance~\cite{Cardoso:2018tly} are shown in red. The red crosses correspond to a point excluded by 5 $\sigma$ whose spectrum is shown in Fig.\ \ref{fig:events}. Bottom: preferred regions of $\sin^2\theta_{23}$-$\delta_{CP}$ for NH (left panel) and IH (right panel) respectively. IH is preferred over NH, with a $\chi^2$ difference of 2.01 at the respective minima.
}
\label{fig:gpres}
\end{figure*}

In this work, we re-examine the effect of $L_\mu-L_\tau$ dark photons on long baseline oscillations, incorporating as well data from the MINOS experiment \cite{MINOS:2011neo,MINOS:2020llm}, and with attention to both possible mass orderings. We find limits
on $g'$ versus $m_{A'}$ that improve upon past analyses~\cite{Brdar:2017kbt,Alonso-Alvarez:2023tii}, and that extend to higher $m_{A'}$ values than other studies of these datasets~\cite{Lin:2023xyk}.
%are compatible with those of Ref.\ \cite{Lin:2023xyk}, extending them to higher $m_{A'}$ values.
%\GA{I rephrased this since we disagree with Lin:2023xyk}
In agreement with previous studies, including a recent combined analysis by the NO$\nu$A and T2K collaborations \cite{combined},
we find that the inverse mass ordering is preferred for reconciling the mild tension between the two experiments, while the new physics effects give a statistically insignificant improvement.

 \section{Methodology}
\label{sec:methodology}

We modify the 
General Long Baseline Experiment Simulator
(GLoBES) \cite{Huber:2004ka,Huber:2007ji}
to use the oscillation probabilities for
$\nu_\mu$ disappearance and $\nu_e$ appearance from beams which are initially
$\nu_\mu$, coming from the Hamiltonian (\ref{Heq}).  This requires numerically solving the Schr\"odinger equation with initial condition $\psi_\mu = (0,1,0)^T$, in the presence of the time-varying $g'A'$ contribution; then the disappearance probability at a distance $L$ is given by
$P_d = 1 - |\langle \psi(L)|\psi_\mu\rangle|^2$, in natural units $c=1$, while the appearance probability is $P_a = |\langle \psi(L)|\psi_e\rangle|^2$.  

As the data is obtained over a time scale much larger than the dark matter oscillation period $1/m_{A'}$, the value of $\alpha$ can be assumed to be independent for each event and uniformly distributed. We therefore average the transition probabilities over $\alpha$. Furthermore, the angle $\phi$  between $\vec A'$ and $\vec p_\nu$ is also expected to vary daily due to the Earth's rotation, and one should therefore also average over it. This variation, however, is much more challenging to model as it depends on the relative orientation of the earth's rotation axis, the neutrino beam, and the dark photon's polarization, which is unknown. To simplify the analysis, we will follow the common practice in the literature and assume $\cos\phi\sim 1$, leaving more detailed exploration of these geometrical effects to future studies. Because the oscillation period of $\phi$ is much larger than that of the neutrino flavor oscillations, $\phi$ can be treated as a constant in the Schr\"odinger equation. To account for values $\cos\phi < 1$,  one should rescale $g'\to \langle |\cos\phi| \rangle g'$ in Fig.\ \ref{fig:gpres}, with the average taken over the experiment's duration.  (A detailed study of the possible time dependence of 
$\phi$ would require knowing the exact times during which experimental data were taken.)
Hence, there are only two new physics parameters to be considered: $g'A'$ and $m_{A'}$.

We combine the neutrino oscillation data from MINOS \cite{MINOS:2011neo}, \nova \cite{NOvA:2021nfi,NOvA:2023iam} and T2K \cite{T2K:2023smv}, fixing the solar parameters at $\theta_{12}=33.41\degree$ and $\Delta m_{21}^2=7.41\times 10^{-5}\ \mathrm{eV}^2$ \cite{Esteban:2020cvm,nufit}. $\theta_{13}$ and $\Delta m_{31}^2$ are varied within their 3-sigma region, which is respectively $[8.19\degree,8.89\degree]$ and $[2.428,2.597]\times 10^{-3}\ \mathrm{eV}^2$ for NH and $[8.23\degree,8.90\degree]$ and $[-2.513,-2.328]\times 10^{-3}\ \mathrm{eV}^2$ for IH. The other parameters $\theta_{23}$, $\delta_{CP}$, $g'$ and $m_{A'}$ are varied over the ranges shown in Fig.\ \ref{fig:gpres}. We assume a prior on $\theta_{13}$ with $\sin^2\theta_{13}=0.02203\pm0.00057$ for NH and $\sin^2\theta_{13}=0.02219\pm0.00058$ for IH.

\section{Results}
\label{sec:results}

Since we are most interested in the case that the $A'$ field comprises the dark matter of the Universe, we fix the value of $A'$ to that reproducing the observed dark matter abundance.
Thus, $A'$ is determined by $m_{A'}$, since the average energy density is given by
$\rho = \sfrac12 m_{A'}^2 A'^2$.  One must distinguish between the cosmological value
$A'_U$ and the local value $A'$, since dark matter is clustered within the galaxy and is therefore more dense on Earth than in the Universe on average.  Following Ref.\ \cite{Alonso-Alvarez:2023tii}, we take 
$A' = 25\,{\rm MeV}\times(10^{-10}\,{\rm eV}/m_{A'})$.
This allows to constrain the gauge coupling $g'$ as a function of $m_{A'}$.

Before presenting our analysis, we  emphasize that this model does not resolve the tension between T2K and NO$\nu$A, as can be seen from Fig.~\ref{fig:prob}. To calculate the oscillation probabilities, we have fixed the standard and nonstandard parameter values at the best-fit values coming from the combined analysis of NO$\nu$A, T2K, and MINOS data. One sees that the oscillation probabilities for the $\nu_e$ appearance channel, which is mostly responsible for the tension \cite{Rahaman:2021zzm, Rahaman:2022rfp}, as well as the $\bar{\nu}_e$ appearance channel, are unchanged by including  dark photon effects in either experiment. The $\nu_\mu$ and $\bar{\nu}_\mu$ disappearance channel probability also remains unaffected by inclusion of the dark photon. Hence, the tension cannot be resolved and the results for the standard oscillation parameters remain unchanged.

We find best-fit values $g'\sim 4\times 10^{-23}$ and $m_{A'}\sim 10^{-12}$\,eV, but with only a small improvement over the SM by $\Delta \chi^2 = 1.5$ (IH) or 2 (NH), indicated on Fig.\ \ref{fig:gpres}
(top row) by the red dots.   Within the SM by itself, there is mild preference for
the IH from these data, which persists within the DM models at the level
of $\Delta \chi^2 \sim 2$. 
Since the DM model has two additional parameters, these are not statistically
significant, hence we emphasize the upper limit curves for $g'$ versus $m_{A'}$.
The $3\sigma$ upper limit curve can be fit in the region shown by $y = 0.73 + 0.97\, x + 0.19\,x^2 + 0.08\,x^3$, where $y = \log_{10}g' + 24$ and $x = \log_{10}m_{A'}/{\rm eV} +13$. Outside of this region, 
the limit on $g'$ can be expressed as a broken power law,
\be
    g' \lesssim \left\{\begin{array}{cc}
        6.3\times 10^{-11}\left(m_{A'}/ {\rm eV}\right),& m_{A'} \ll 10^{-12}{\rm\,eV}\\
        100 \left(m_{A'}/ {\rm eV}\right)^2,& m_{A'} \gg 10^{-12}{\rm\,eV}
\end{array}\right.\,,
\ee
whose analytic form has been explained in Refs.\ \cite{Alonso-Alvarez:2021pgy,Alonso-Alvarez:2023tii}.
The cross-over occurs for dark photon oscillation frequencies that are of the same order as the neutrino oscillation frequency, $m_{A'} \sim \Delta m^2_{23}/4E$.
Analytic approximations can be used for large and small $m_{A'}$ to avoid having to solve the Schr\"odinger equation numerically in these regimes.

\begin{figure*}[t]
\centerline{\includegraphics[width=0.67\columnwidth]{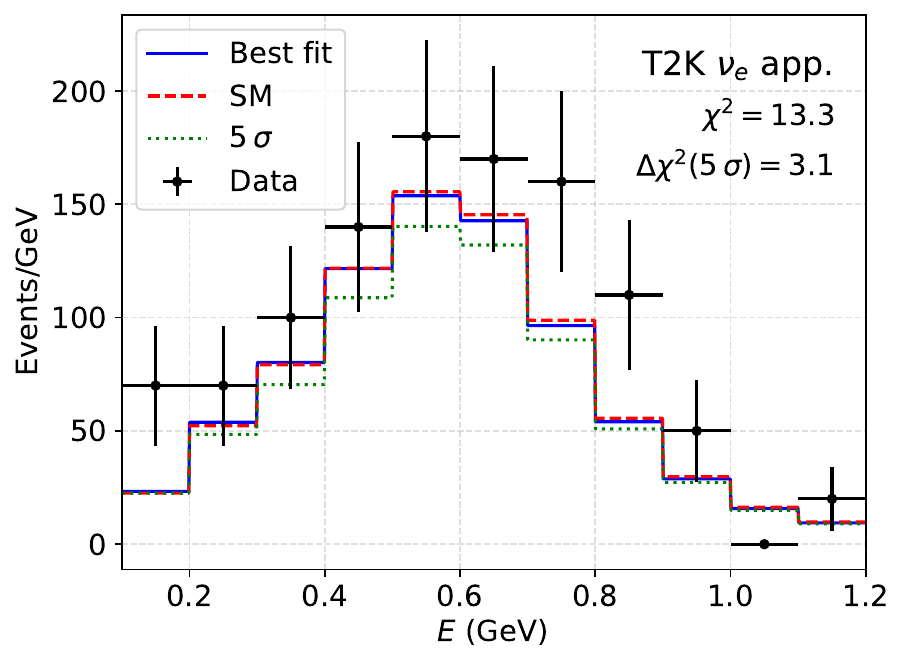}
\includegraphics[width=0.67\columnwidth]{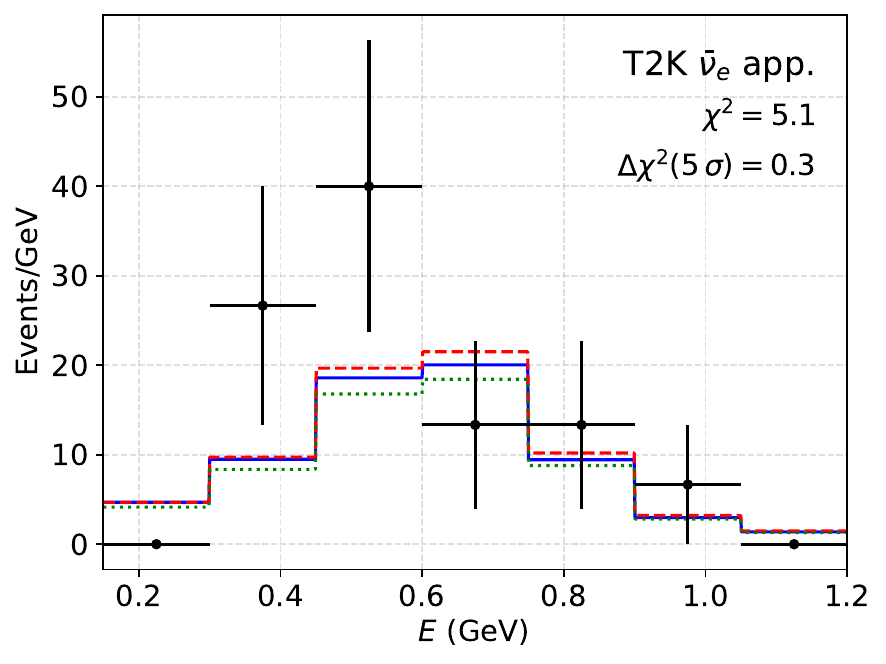}
\includegraphics[width=0.67\columnwidth]{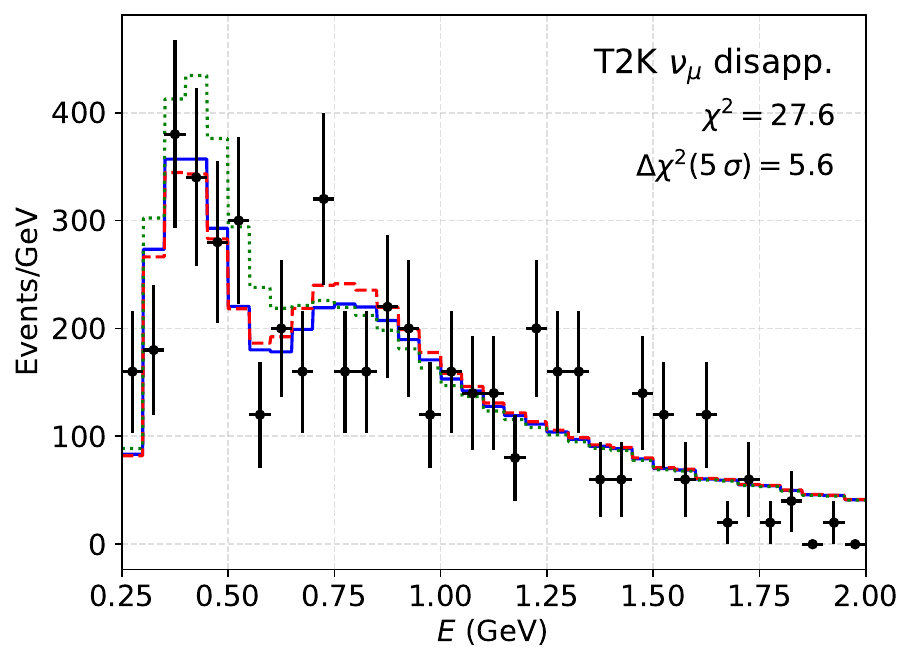}}
\centerline{\includegraphics[width=0.67\columnwidth]{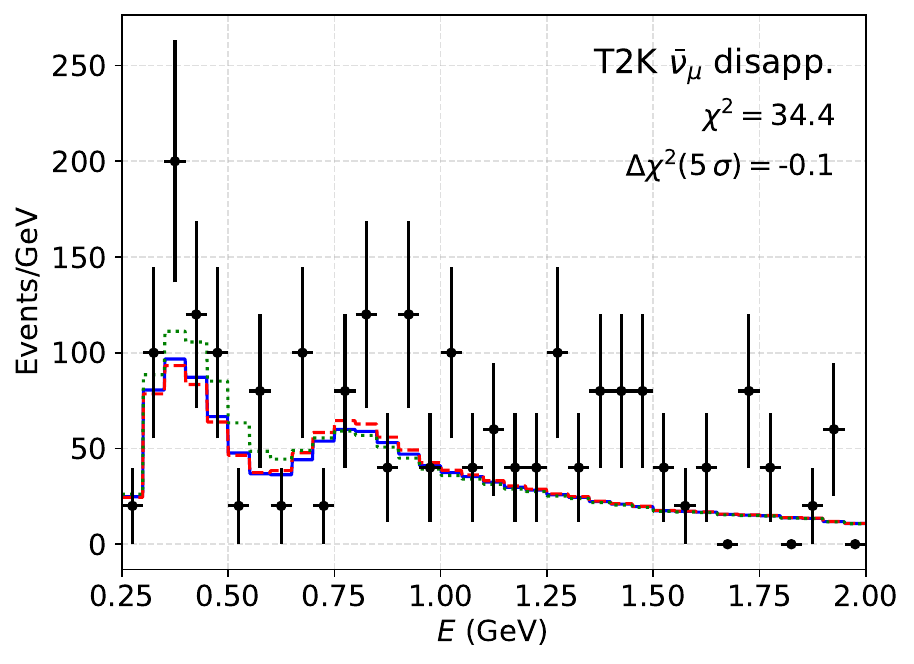}
\includegraphics[width=0.67\columnwidth]{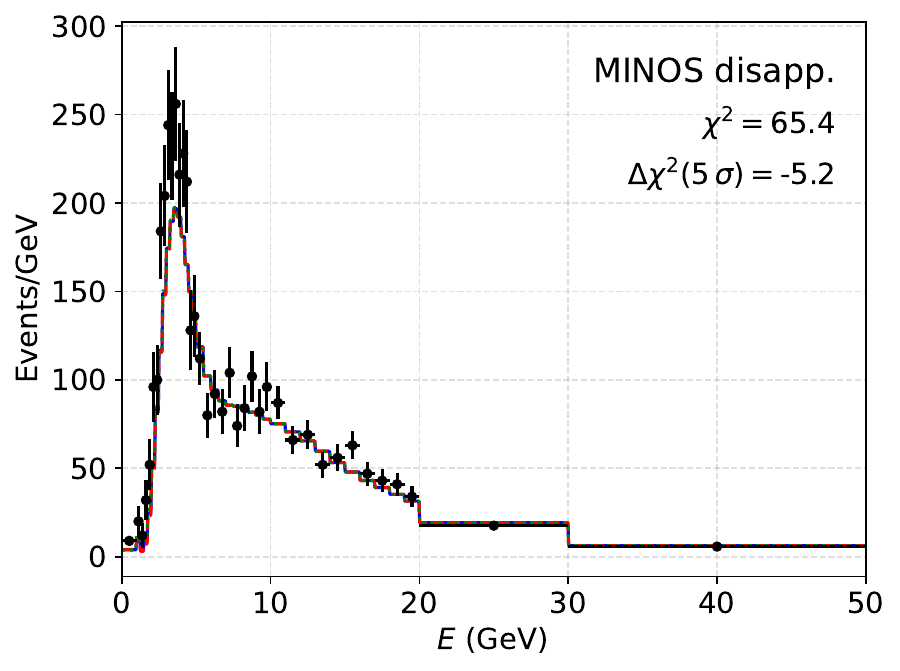}
\includegraphics[width=0.67\columnwidth]{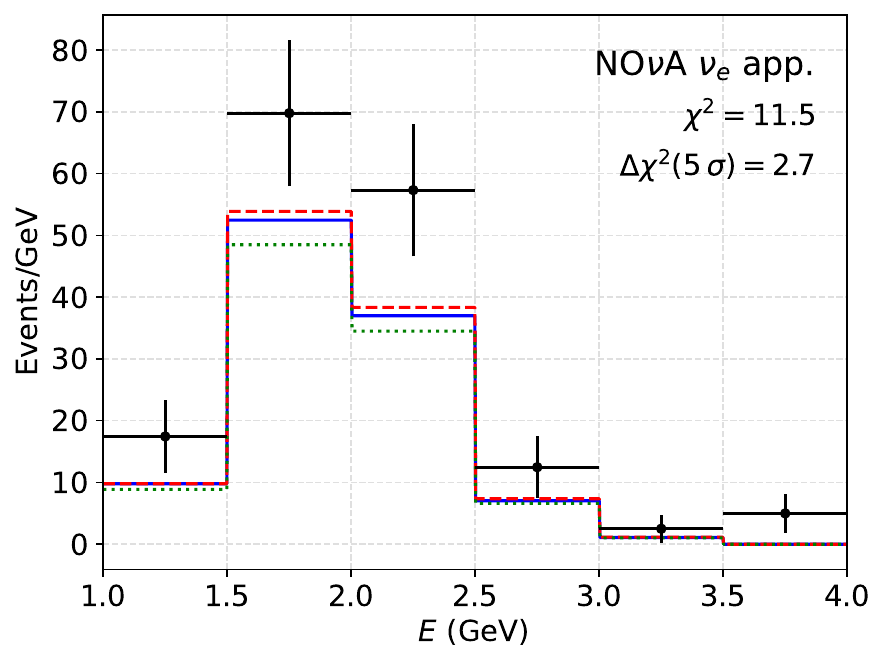}}
\centerline{\includegraphics[width=0.67\columnwidth]{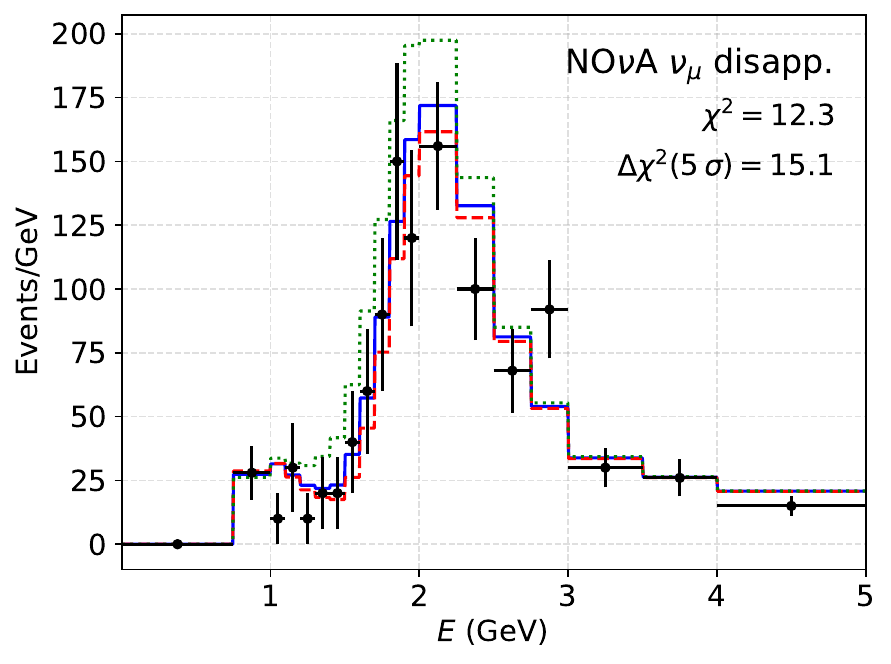}
\includegraphics[width=0.67\columnwidth]{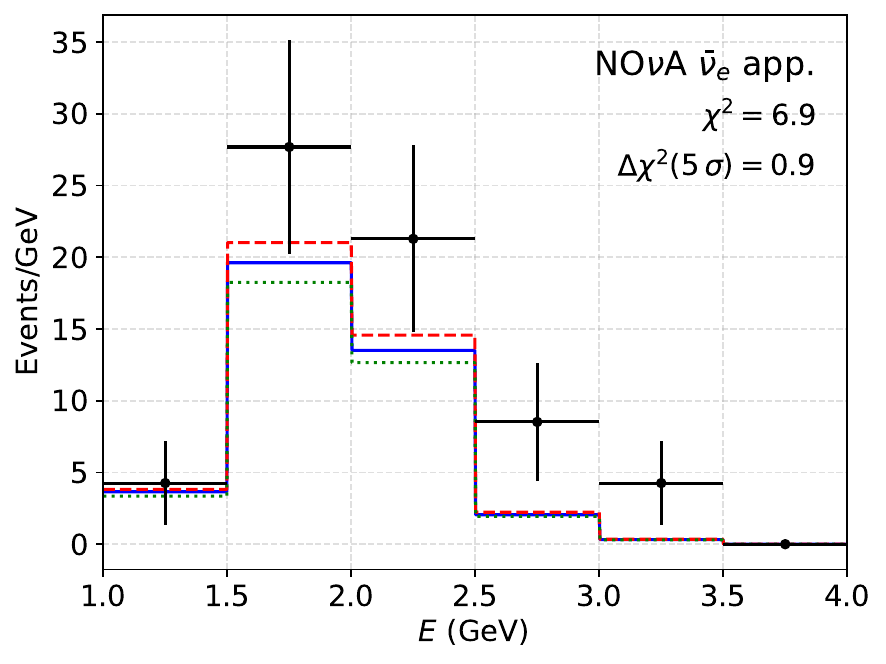}
\includegraphics[width=0.67\columnwidth]{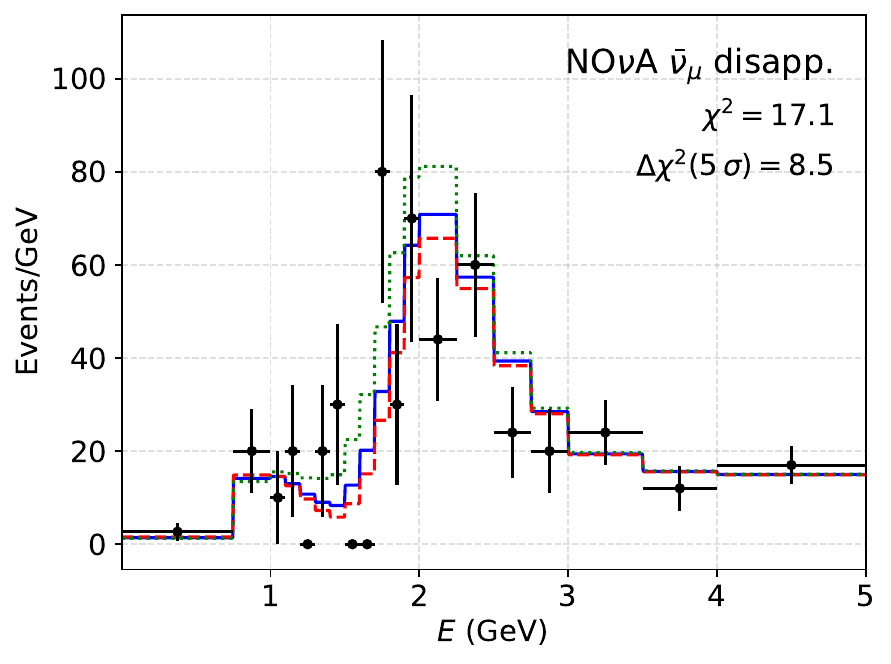}}
%\vspace{}
\caption{Observed and predicted number of events at T2K, NO$\nu$A and MINOS. The predictions were obtained with the best-fit values of our combined analysis of the three experiments. The green dotted lines correspond to the point shown by a red cross in Fig.\ \ref{fig:gpres}, which is excluded by 5 $\sigma$. We also show the $\chi^2$ value for the best fit of each channel and the $\Delta\chi^2$ between the 5-$\sigma$ point and the best fit.
}
\label{fig:events}
\end{figure*}

Comparing to previous results, we obtain a factor of $\sim 5$ stronger limit than Refs.\ \cite{Brzeminski:2022rkf,Alonso-Alvarez:2023tii}, which used the two-flavor approximation and only T2K data (taken prior to Ref.\ \cite{T2K:2023smv}).
The improvement in the constraints is driven by the larger dataset that we are using, which contains roughly ten times more events than previous analyses from nine distinct channels.
As can be seen in Fig.~\ref{fig:events}, the inclusion of NO$\nu$A data in the analysis plays a particularly important role in strengthening the constraints on the dark photon parameters.
Furthermore, the addition of the MINOS data helps to break degeneracies between the new physics and the neutrino vacuum mixing parameters. 
In contrast, Ref.\ \cite{Lin:2023xyk} obtains an ostensibly much stronger limit on $g'$, by a factor of $\sim 2500$.  Based on discussion with the authors, we believe this was due to a numerical error.

The bottom row of Fig.\ \ref{fig:gpres} shows the confidence intervals on the
$\sin^2\theta_{23}$-$\delta_{\sss CP}$ plane in the two assumed mass hierarchies. 
On the NH plane, the tension between T2K and NO$\nu$A cannot be reduced by the neutrino-dark photon interaction.
This can be confirmed by comparing the joint analysis to separate fits to the individual experiments.
As explained in ref.~\cite{Rahaman:2021zzm}, the tension arises from the $\nu_\mu \to \nu_e$ appearance channel. Since our neutrino-dark photon interaction model does not have similar kind of impact on the $\nu_\mu \to \nu_e$ oscillation probability, the tension at NH plane does not get reduced.
The combined analysis of NO$\nu$A and T2K, along with MINOS, prefers IH over NH. For IH, the individual experiments as well as their combined data prefer $\delta_{\sss CP}\sim-90^\circ$. 
For NH, the allowed region on $\sin^2\theta_{23}$-$\delta_{\sss CP}$ for the combined analysis is closer to the T2K allowed region, because of the larger statistics of T2K.

In Fig.~\ref{fig:events}, we have shown the expected event spectrum at the best-fit point of our analysis, as well as the best-fit point in the standard 3-flavor oscillation analysis, along with the observed data points for all the three experiments and for all the different channels considered in this paper. As can be seen from the figure, the expected event spectrum at the best-fit point for both the cases of dark photon and standard case do not differ significantly. 
\section{Summary and conclusions}
\label{sec:summary}

  In Ref.\ \cite{Alonso-Alvarez:2023tii} it was shown (in the two-flavor approximation) that for $m_{A'}$ above the critical value $m_{A'}\sim \Delta m_{23}^2/4E$ discussed above, 
$A'$ oscillations can be integrated out, and their effect on the $\nu$ oscillations is described by a shift in the effective
$\Delta m^2_{23}$, 
\be
    [\Delta m^2_{23}]_{\rm eff} = 
    \Delta m^2_{23}\left(1 -\left(g'A'\sin2\theta_{23}\over 2 m_{A'}\right)^2 \right).
    \label{dm2eff}
\ee
This has the consequence that the true mass splitting $\Delta m^2_{23}$
is larger than the effective value inferred in the $L_\mu-L_\tau$ dark matter background. Eq.\~(\ref{dm2eff}) thus describes a degeneracy between the vacuum mixing parameters and the new physics parameters in the region $m_{A'}\gg 10^{-12}\,$eV.
This could lead to an interesting interplay between the mass splittings determined in local oscillation experiments and cosmological bounds on the sum of neutrino masses~\cite{DESI:2024mwx,Wang:2024hen,Allali:2024aiv}, in a similar way to the models proposed in~\cite{Sen:2023uga,Sen:2024pgb}, which merits a separate study.

We remind the reader that $A'$ in Eq.\ (\ref{dm2eff}) is shorthand for the component of $\vec A'$ parallel to the neutrino beam.  Thus $A'$ will vary by a factor depending on the angular orientation of the baseline relative to $\vec A'$, which differs between various experiments, and which varies with time due to the motion of the Earth. This effect was partially considered in Ref.~\cite{Brzeminski:2022rkf}, and we hope to investigate it in greater detail in the future along with the secular dynamics arising from a dark photon polarization varying at the coherence timescale.

%This could help alleviate the recently arising tension between the mass splittings determined in local oscillation experiments and recent cosmological bounds on the sum of neutrino masses~\cite{DESI:2024mwx,Wang:2024hen,Allali:2024aiv}, in a similar way to what is proposed in~\cite{Sen:2023uga,Sen:2024pgb}.
%\GA{I changed the discussion here a bit since I think it is an interesting possibility.}
%Current bounds from cosmology \cite{DiValentino:2024xsv} are tending to leave little room for a discrepancy between $\Delta m^2_{23}$ and $[\Delta m^2_{23}]_{\rm eff}$, so we do not further pursue this possibility here.

\section*{Acknowledgements}

GA and JC are supported by the Natural Sciences 
and Engineering Research Council (NSERC) of Canada. BL is supported by the Fonds de recherche du Québec Nature et technologies (FRQNT). JC and BL thank the CERN
Theoretical Physics Department for its generous hospitality while this work was being completed.
We thank J.\ Kopp for discussions and kind assistance with incorporating MINOS data within GLoBES, and H.-X.\ Lin, J.\ Tang, and S.\ Vihonen for discussions.

\bibliography{ref}
\bibliographystyle{utphys}

\end{document}